\documentclass[10pt,conference,compsocconf]{IEEEtran}
\usepackage{times}
\usepackage{mathptmx}
\usepackage{graphicx}
\usepackage{times}
\usepackage{array}
\usepackage{adjustbox}
\usepackage{lipsum}
\usepackage{comment}

\usepackage{caption}
\ifCLASSINFOpdf
\else
\fi
\begin{document}
\pagenumbering{gobble}
%
\title{\textbf{\Large A Study on 3D Surface Graph Representations} \vspace{-2ex}}



%

\author{\IEEEauthorblockN
{
Long H. Nguyen,
Abdullah Karim
}
\IEEEauthorblockA{Texas Tech University, Lubbock, Texas, USA}
long.nguyen@ttu.edu,
abdullah.karim@ttu.edu
}


\maketitle


\begin{abstract}
Surface graphs have been used in many application domains to represent three-dimensional (3D) data. Another approach to representing 3D data is making projections onto two-dimensional (2D) graphs. This approach will result in multiple displays, which is time-consuming in switching between different screens for a different perspective. In this work, we study the performance of 3D version of popular 2D visualization techniques for time series: horizon graph, small multiple, and simple line graph. We explore discrimination tasks with respect to each visualization technique that requires simultaneous representations. We demonstrate our study by visualizing saturated thickness of the Ogallala aquifer - the Southern High Plains Aquifer of Texas in multiple years. For the evaluation, we design comparison and discrimination tasks and automatically record result performed by a group of students at a university. Our results show that 3D small multiples perform well with stable accuracy over numbers of occurrences. On the other hand, shared-space visualization within a single 3D coordinate system is more efficient with small number of simultaneous graphs. 3D horizon graph loses its competence in the 3D coordinate system with the lowest accuracy comparing to other techniques. Our demonstration of 3D spatial-temporal is also presented on the Southern High Plains Aquifer of Texas from 2010 to 2016.
\end{abstract}



%
\IEEEpeerreviewmaketitle

\section{Introduction}

Time series charts are graphical representations of quantitative values changing over time ~\cite{timeseries}. Line chart is the simplest way to visualize time series where we have one axis represents for a time period and the other axis represents for its quantitative values. Since its invention by William Playfair (1875-1923)~\cite{wikiWilliam}, a lot of research has been conducted and has invented different variants of time series with better visualization in certain ways. One approach that researchers try to target is the representation of multiple time series where line charts visualization become difficult because of the limited vertical space, as well as its high visual clutter. Two other popular techniques are small multiples and horizon graphs that we will discuss later in this paper.

Small multiples represent lines in its own space to reduce visual clutter. Because space is allocated per line, it requires to compact itself in order to fit inside the vertical screen resolution. This causes difficulty in value comparing between points on the same line due to the scale reduction. Later on, horizon graph was invented. In horizon graph, the gap between the line and baseline is chunked into multiple layers, and the higher layer is overlaid on top of lower ones. The color is also used to fill the gap between baseline and the area formed by the line chart and the boundaries of its layer. The horizon graph is proven as space reduction while gaining identity and information extraction solution for multiple simultaneous time series representations in 2D visualization ~\cite{reijner2008development}. This solves the problem of visual clutter while still representing multiple series.

Horizon graph which is represented in 2D space has grown its usage since its first invention. In recent years, people are increasing to higher demand with 3D or even multi-dimensional data expecting multi-dimensional space or even merging to virtual space to see the representation. Together with this trend, 3D visualization techniques also evolve to solve more complicated information visualization demand from multi-dimensional data. With a standard surface graph, users can deal with few simultaneous series only.
 
In this work, we study 3D version of split space techniques (small multiples and horizon graphs) to accommodate space and data comparison problems in the world of 3D visualization. We address the lack of perception capabilities in representing 3D data by evaluating various tasks requiring multiple concurrent displays through monitored laboratory experiments. Our main motivation for this research is to provide guidelines for graphical designers who are in need of methods to represent 3D spatial-temporal graphs for their applications. In addition, we stress advantages and disadvantages of each graph type for certain use cases. In the demo, we represent spatial-temporal data visualization of the saturated thickness of the Ogallala aquifer. In summary, our main contributions are:

\begin{itemize}
\item We present 3D version of horizon graph and small multiple that can address the space and perception capability in presenting multiple concurrent series.
\item We provide a guideline for 3D designers in visualizing surface graph data.
\item We demonstrate our sample visualizations in the Southern High Plains Aquifer of Texas and conduct a user study for each visualization technique.
\end{itemize} 

We understand that providing more interactions will bring more capabilities to users in catching information from the graph. However, to evaluate the visual effect itself, we will just provide basic interaction such as zooming, and rotating graph meshes in the 3D coordinate system. Other alternative methods such as hierarchical aggregation ~\cite{elmqvist2010hierarchical}, temporal queries ~\cite{hochheiser2004dynamic}, temporal clustering ~\cite{kaufman2009finding} or further interactions are not considered in the scope of this paper. 

Even though our demonstration is the spatial-temporal data of the Ogallala aquifer, we believe that the technique and its analysis result can be applied to any multi-dimensional data visualization representing in 3D coordinate system.

The rest of this paper is organized as follows: We discuss the related work in the next section. Then we provide an overview of our research questions, our evaluation criteria and discuss some current 3D visualization techniques in representing surface graphs. In Section ~\ref{sec:user-study}, we discuss our study in detail with hypothesis, equipment, tasks and study design. We analyze and discuss our result in Section ~\ref{sec:results}. Finally, we conclude our paper with future plans. 

\section{Related Work}
In this section, we do not intend to survey all visualization and perception techniques for geospatial-temporal representations ~\cite{bach2014review}. Instead, we will discuss some related research on graphical perception and some spatial-temporal visualization techniques in two or 3D coordinate systems.

\subsection{Graphical perception}
Graphical representation of statistical data evolved for years even before computers were invented. The initiative work presented by Eells et al. at ~\cite{eells1926relative} starting toward a higher demand which different graphical representations are compared to obtain a greater visual effect. In the next year, Croxton et al.~\cite{croxton1927bar} compared bar charts to circle diagrams for the accuracy of judgment. He then continued to question why most users of statistical charts found linear comparison represented by charts has more accuracy than other area charts comparison. He discussed the related capability of bars, squares, circles, and cubes by some simple comparisons at ~\cite{croxton1932graphic}. Peterson et al. at ~\cite{peterson1954accurately} tried to figure out ``How accurately are different kinds of graphs read?'' by reading values from different graphical representations of statistical data. 

Cleveland et al. at ~\cite{cleveland1984graphical} set scientific foundation for graphical analysis and data representation under the name ~\textit{graphical perception} and defined it as the visual decoding of information encoded on graphs.  Simkin et al. at ~\cite{simkin1987information} confirmed by evidence that people have generic expectation about types of information will be the major messages in some types of graph. A comparison is conducted and revealed that the most accurate of judgment is in bar charts then divided bar charts and least accurate is the pie charts. However, the representation involves only two charts simultaneously. 

Lohse et al. at ~\cite{lohse1991cognitive} presented a cognitive model to understand perceptual and cognitive processes people use to decode information from a graph. They conducted an empirical study in predicting reaction time, level of difficulty to acquire information between computer simulated graphical perception and actual performance of users. After two years, Gillan et al. ~\cite{gillan1993componential} discussed a componential model of human interaction with the graphical display. From human-computer interaction point of view, the model explains human interaction with graphs. Later on, he brought this model to study with various graphical representations, such as line graphs in linear modeling, scatter plots, stacked bars, and pie charts.
 
Huang et al. ~\cite{huang2008beyond} argued that time and error measures are limited in providing essential knowledge that is useful for graphical design. They conducted three user studies to evaluate and demonstrate the usefulness of cognitive perspective that goes beyond merely measuring time and error. Spence et al. ~\cite{spence1991displaying} used a series of experiments to investigate the perception of percentages in bar charts, pie charts, and tables. And very recently, Heer et al. ~\cite{heer2009sizing} conducted two controlled experiments to measure the effect of chart size and layering and evaluate their effect in value comparison tasks.

Barfield et al. at ~\cite{barfield1989effects} discussed the relationship between 2D or 3D graphs displayed on paper or computer and the problem-solving performance of experienced and novice manager with respect to each representation by measuring solution times, confidence in answers and effectiveness of solutions.

Each of these studies are focus on how a visual encoding and representing affects the accuracy and/or response time of discriminating values of the underlying data. Nevertheless, most of these researches discussed those visual factors with the 2D visualization techniques. There are still remaining questions on how well these techniques perform when dealing with multi-dimensional data and representing as 3D charts.

\subsection{Spatial-temporal data visualization}
One of the earliest demonstrations related to spatial temporal data is Napoleon's march towards Moscow. Its visual encoding uses color, size, text annotation, and position to represent different states of Napoleon's army in 1812~\cite{Tufte1986VDQ} in one snapshot. Nowadays, data visualization tools become more important when the demand in data dimension increased. David et al. ~\cite{david2015web} presented abstract interface using recent web based JavaScript libraries to visualize large complex sets of spatial temporal data over the web.

D{\"u}bel et al. ~\cite{dubel20142d} shown that 2D and 3D visualization techniques exhibit different advantages and disadvantages in human perception. Aigner et al. ~\cite{aigner2011visualization} discussed analytical view on factors, and providing samples to visualize time oriented data. Andrienko et al. ~\cite{andrienko2004interactive} introduced the use of interactive visualization tools to address problems of spatial temporal data. 

Tominski et al. ~\cite{tominski20053d} brought 3D icons into a map display for representing spatial-temporal data. There are also event-based methods being integrated for reducing the amount of
information to be represented. The approach relies on two popular concepts: 3D information visualization and information hiding. The technique is inherited by three-step process for information visualization: overview first, zoom and filter, and then details-on-demand which were presented by Keim et al.~\cite{Keim2004Infor-5420}.

In supporting with the third dimension, Carneiro ~\cite{carneiro2008communication} presented a case study of Geneva with 3D Geneva project. It aims to evaluate if users would
potentially be interested to integrate the third dimension in visualizing the available geographical data. Koussa et al. ~\cite{koussaproposal} proposed web based solution for 3D geographical information system using multilevel spatial database structure and layer management technique to visualize and analyze the data. 

Treinish ~\cite{treinish2000visual} discussed human problem solving and decision making performance varies enormously (100:1) with different presentations of high-resolution incorporated predictive weather data. More recent work was done by Kehrer et al. ~\cite{kehrer2013visualization} who summarized existing methods for visualization and interactive visual analysis of multifaceted scientific data and proposed a categorization of approaches. 
\section{Definitions, Data Set and Methodology}
In this section, we discuss some common definitions that are used throughout the paper. We also briefly introduce our data set and methodology in this research.

\subsection{Definitions}
\textbf{Saturated thickness} is the vertical thickness of a hydro-geologically defined aquifer in which the pore spaces of the rock forming the aquifer are filled with water~\cite{Schloss2000current}.

\textbf{Graph type} is the type of graph used in representing information. In this research, graph type is either surface graphs, small multiples or horizon graphs. If we do not mention whether these types are 3D or 2D, it is implicitly understood that the graphs are in 3D coordinate system.

\textbf{Study year} is the year that the study data belongs to. When discussing its graphical representation, we mean that it is a surface graph representing the data of that year.

\subsection{Data Set}
We use saturated thickness data from 2010 to 2016 to create visualizations in the study. We choose this data set because it has both spatial, temporal dimensions which fulfill our research objective in this work. Besides, these data sets are part of our research project in visualizing the aquifer.

\subsection{Methodology}
Our research goal is to effectively visualize the saturated thickness of the Ogallala aquifer in a 3D coordinate system that requires little space but also maintains the visual richness necessary for comparison tasks. To do so, we first designed a model in which  the geographical locations encompassing the Ogallala were used as the base ground. Because this model considered only the thickness of the Ogallala at a given location, these locations described a flat plane. The surface of the model represented the saturated thickness of the Ogallala, producing an image resembling a terrain and so called a surface graph. We represent the surface graph via different techniques (simple surface graph, small multiple and horizon graph) and conduct a user study to evaluate its effectiveness based on user perception. 

\section{Evaluation Criteria and Visualization Techniques}
In this section, we discuss evaluation criteria and visualization techniques individually to justify each case.

\subsection{Evaluation Criteria}
We reuse some evaluation criteria described in~\cite{javed2010graphical}  to evaluate each surface representation. Table ~\ref{table:graph-type-property} summarizes attributes for the three surface techniques surveyed in this paper. We discuss these techniques in the following section.

\begin{itemize}
\item \textbf{Space management}: This criterion describes how space is utilized in a graphical representation. In other words, it is whether space is shared or split into multiple occurrences. Shared space is represented in the same coordinate system. So, all series are overlaid; hence, it is easier for comparison. The height of the space required is proportional to the biggest quantitative value of the data in all study years. However, if there are more study years, it may introduce clutter which makes it hard for users to identify the graphs and compare one to another. On the other hand, in split space visualization, whether there are less or more occurrences, each graph has its own space to be presented. It means that this technique requires the space to be compressed to fit all the graphs. Hence the perception of information may be reduced.  

\item \textbf{Space per study year}: This is the vertical and horizontal (a rectangular cube) amount of display space allocated to each graph. This characteristic is not a key factor, but it is important as any visualization should fit in a monitor screen size. We aim to use as little space as possible without negative consequences to its visualization effect.

\item \textbf{Identity}: This criterion tells how easy it is to distinguish between occurrences. Obviously, it is more difficult to identify graphs in shared space visualization. It is required to use graphical methods such as colors and styles to convey the graph identity. The split space techniques preserve the identity of the graphs.

\item \textbf{Visual clutter}: This factor refers to clutter associated with representation techniques for a large number of graphs.

\end{itemize} 

In addition, we add a new criteria to evaluate the effectiveness of the representation technique, called \textit{information identification}. 
\begin{itemize}
\item \textbf{Information identification}: One of the key factors that affects user preference for one graph type over others is its information identification property. One graph stands out from other graphs because it conveys information clearly and makes it easy to find information within the graph. To measure this factor, we can use its accuracy test and the time it takes to complete related tasks.
\end{itemize} 

\begin{table}[h!]
\centering

\caption{PROPERTIES OF 3D VISUALIZATION TECHNIQUES SURVEYED IN THIS PAPER.}
\begin{tabular}{|l|c|c|c|c|} 
\hline
 Visualization & Space & Space per & Visual & Information \\
  Technique & Management & Study Year & Clutter & Identification\\
\hline
 Simple Surface & Shared & S + h & Medium & Medium\\ 

 Small multiple & Split & S/N + h & Low & Medium\\ 

 3D Horizon & Split & S/(N*2*B) + h & Low & Low\\ 
\hline
\end{tabular}
\label{table:graph-type-property}
\end{table}

 In Table ~\ref{table:graph-type-property}, ~\textbf{S} is the available space, ~\textbf{N} is number of study years, ~\textbf{B} is number of bands and ~\textbf{h} is the minimum height required to view the entire surface (to view the depth level of the surface).

\subsection{Simple Surface Graph}
Simple surface graph is a common graph that we represent in 3D space. Each axis represents one dimension of the data. In our demonstration, the ground plane (Oxy) is the longitude and latitude, and the vertical (Z) axis is the saturated thickness. With this type of graph, we can intuitively see that the space required to represent is shared between graphs and equal to the rectangular cube surrounding the vertical height and horizontal width of the graphs.  

\begin{figure}[!h]
 \centering
 \includegraphics[width=1\linewidth]{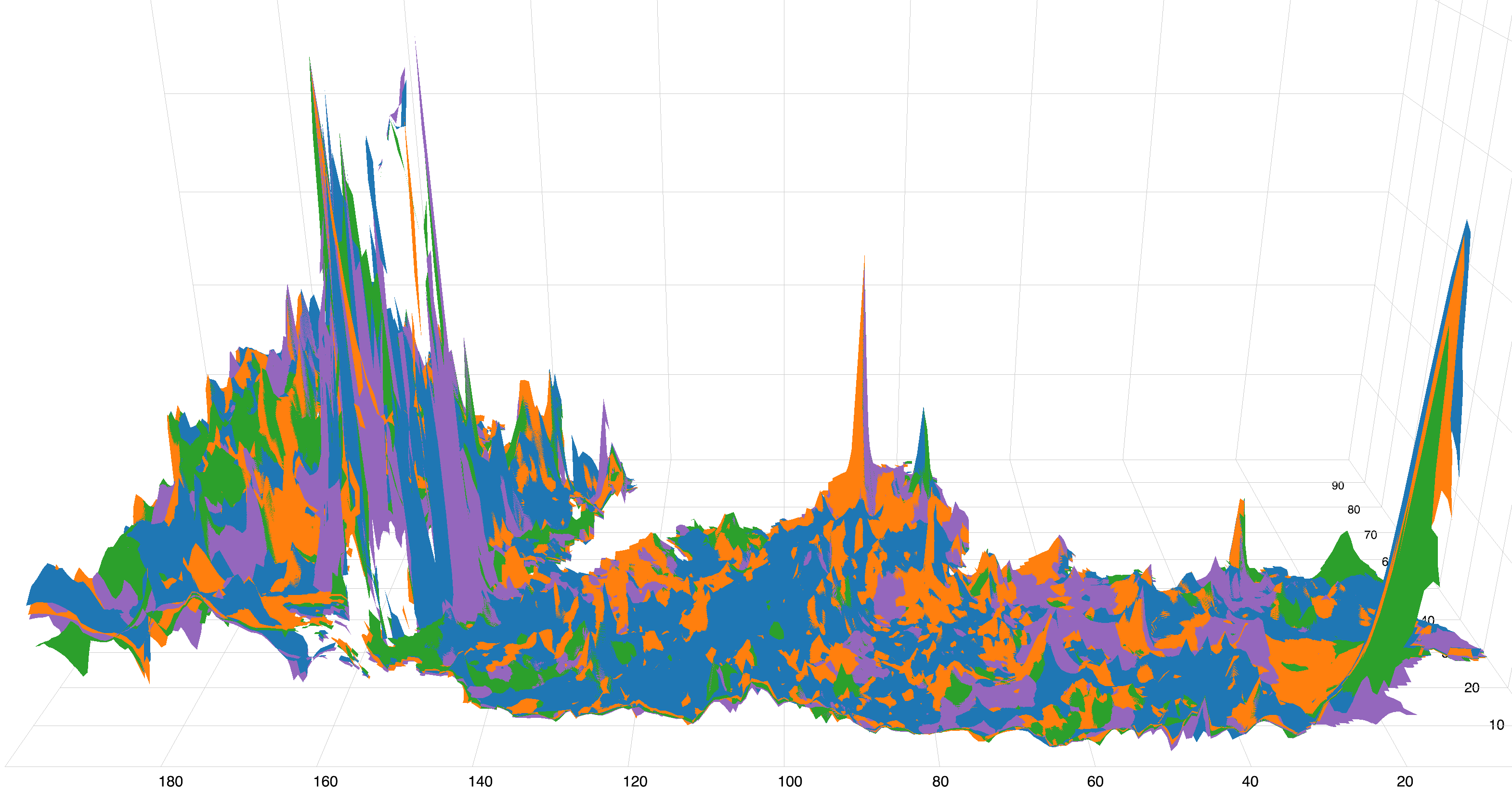}
 \caption{\label{fig:surface-graph} Saturated thickness surface graphs of the Ogallala in 2010 (blue), 2012 (orange), 2014 (green) and 2016 (purple) in the shared space visualization technique.}
 \vspace{-2em}
\end{figure}

Adding more graph surface is achieved by presenting it in the same coordinate system and with different surface colors or styles to identify itself. To make the 3D effect more efficient, we can use the same color with different bands or a line grid along the surface to improve differentiation of height levels in the graph. In this paper, when we overlaid multiple study years, we received informal feedback that multiple colors make confused users, so it is unfair in comparison with other techniques. Therefore, we decided to keep one color for each study year. 

Figure \ref{fig:surface-graph} shows an example with four study years of saturated thickness of the Ogallala. Each study year is represented in a separate color. As shown in Table ~\ref{table:graph-type-property}, these graphs are shared space so that we can easily compare saturated thickness over years. Distinguishing identity is already challenging because the surfaces are overlaid and hide each other. The mix of the colors with the complexity of each graph introduces clutter as well.

\subsection{Small Multiple}
Small multiple for surface graph visualization means that we split the space into sub spaces separated by a plane. Each graph lies inside the allocated space and has its own coordinate system. However, these coordinate systems have the same scale to maintain comparison capability. In other words, with this type of graph, we no longer add graphs into the same graph space hence it is critical to have the same axes scaling across all allocated spaces to allow comparison between graphs. The more we compact the graph, the shorter and smaller amount of space we require to represent them. However, to a certain limit, we can't compress the space more because it will be harder to tell whether it is a 3D mesh or just a thick plane. 

\begin{figure}[!h]
 \centering
 \includegraphics[width=1\linewidth]{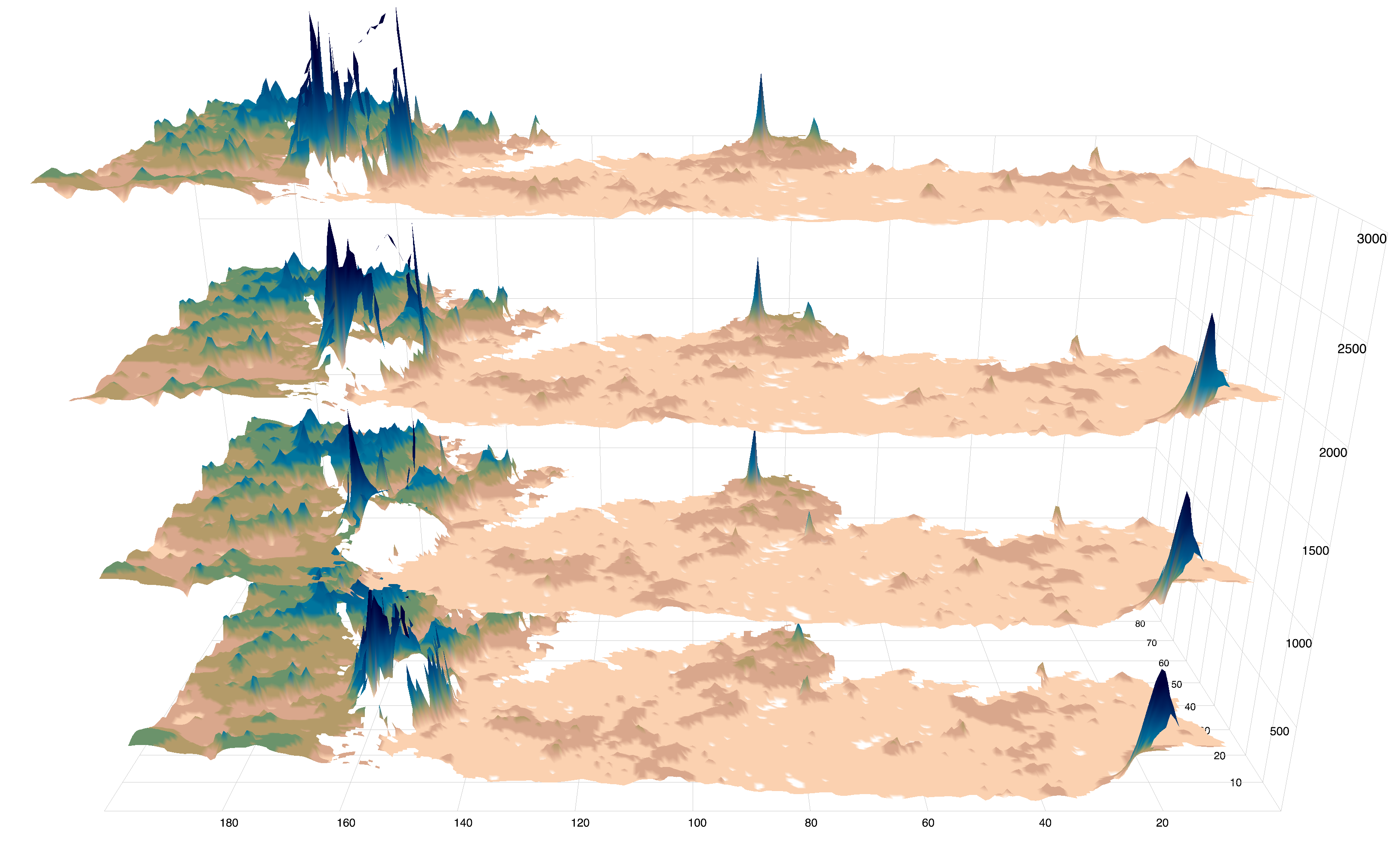}
 \caption{\label{fig:small-multiple-graph} Saturated thickness surface graphs of the Ogallala in 2010, 2012, 2014 and 2016 (from bottom to top) with small multiple representation.}
  \vspace{-1em}
\end{figure}

In Figure \ref{fig:small-multiple-graph}, we demonstrate small multiples of three study years of saturated thickness of the Ogallala aquifer. We rescale the height of the saturated thickness in accordance with the scale of the coordinate system in each allocated space. This made the saturated thickness look shorter compared to its original simple surface graph. Obviously, the space to represent each graph is now horizontally and vertically reduced to \textbf{(S / N) + h} as demonstrated in Table ~\ref{table:graph-type-property}. In addition, we gain the identity property in this case since each year representation is separated. The saturated thickness value can be measured by its height or color represented in the graph. This feature makes 3D small multiple different from its 2D version that the line height is only measured by its value in the Y axis.

\subsection{3D Horizon Graphs}
The horizon graph was originally represented as “two-tone pseudo-coloring” ~\cite{saito2005two}. Latter on, it was developed and introduced under the name "Horizon graph" by the company Panopticon ~\cite{reijner2008development}. The graph is chunked into multiple layers. The upper layer is overlaid on top of a lower layer. Each layer is encoded with a different color to convey its height information. Its construction steps are represented in ~\cite{reijner2008development}.

\begin{figure}[!h]
 \centering
 \includegraphics[width=1\linewidth]{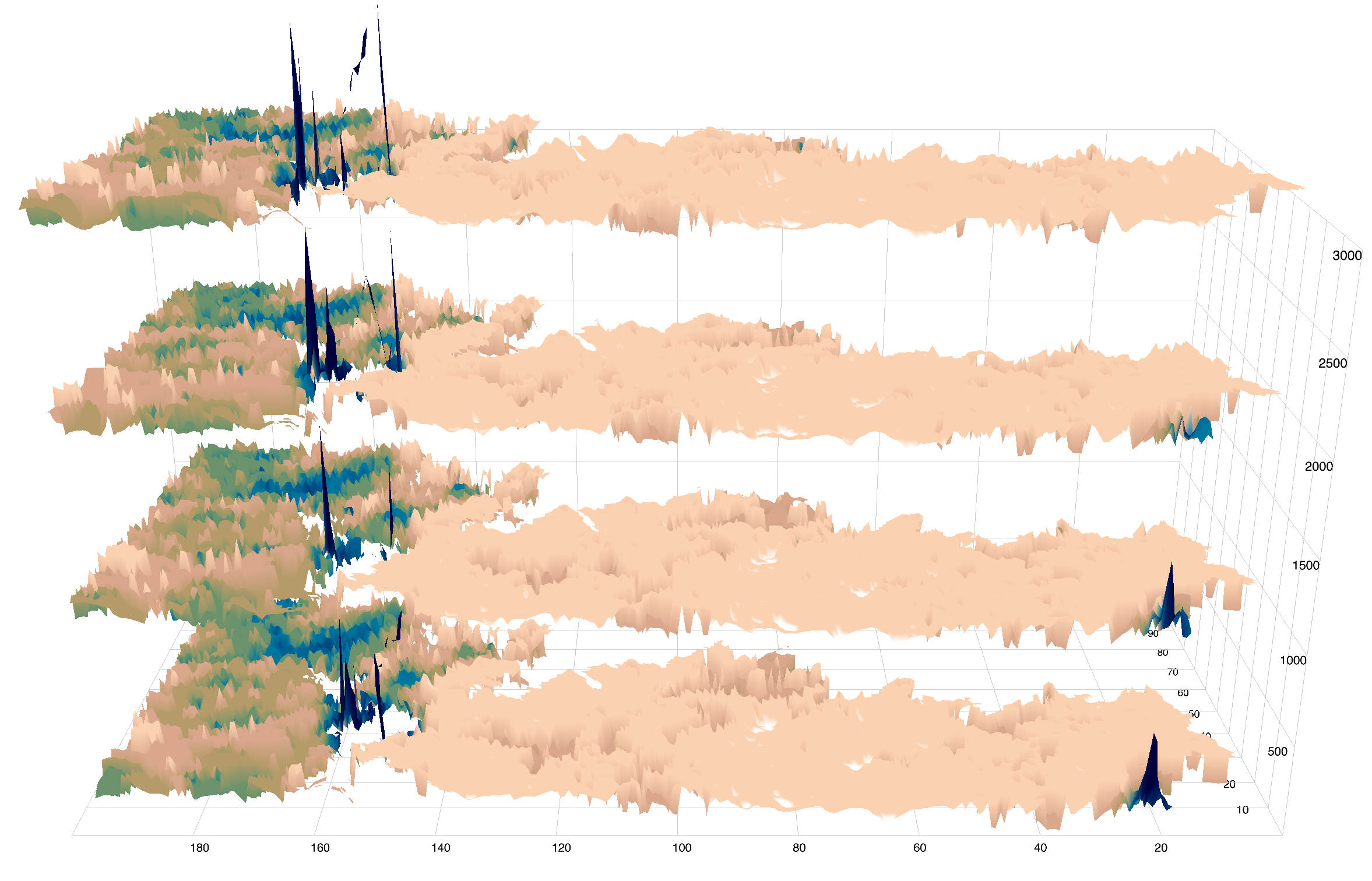}
 \caption{\label{fig:horizon-3d} The 3D horizon graphs of the Saturated thickness of the Ogallala in 2010, 2012, 2014 and 2016 (from bottom to top).}
  \vspace{-2em}
\end{figure}

We reuse the concept of horizon 2D in 3D horizon graph visualization that upper layer of the graph is moved down to the base ground plan. Its color band is still the same as other visualizations. The allocated space per graph is reduced to \textbf{(S / (N * 2 * B)) + h} as described in Table ~\ref{table:graph-type-property}. This makes the horizon graph use the least space compared to other techniques we have discussed.
Figure ~\ref{fig:horizon-3d} provides an example of 3D horizon graphs for four study years.  

\section{User Study}
\label{sec:user-study}
Our intention of this research is to introduce a 3D version of horizon graph and small multiple, and explore user performance on these techniques under different space and cardinality constraints. Does the 3D horizon graph perform better over other techniques? How does each type of graph benefit users in certain conditions? 

To investigate these research questions, we designed a quantitative user study approach to measure correctness performance and time it takes for different combinations of visualization techniques and number of study years. In particular, we present interfaces of surface graphs, small multiples and 3D horizon graphs. With each graph type, we present two, three and four study years for comparison and discrimination tasks. The measurement matrices are correctness of each task's answer and the time requires to complete each trial.

\subsection{Hypothesis}
It is intuitive that shared space surface graph and split space graph techniques have different strengths and weaknesses for tasks with different visual spans. Also the number of study years would strongly affect to user legibility.  

\begin{itemize}
\item ~\textbf{H1} ~\textit{Simple surface graph will outperform other techniques on small number of surfaces}(for example, with 2 surfaces sharing the same coordinate). We predict that, the simple graph will outperform with few study years when the visual clutter is small and has less negative effects. When the number of study years increases, the visual clutter becomes more problematic and deducts graphical perception.
\item ~\textbf{H2} ~\textit{Small vertical resolution will reduce small multiple performance}. We want to confirm this premise that the display space will have a strong impact to user performance.
\item ~\textbf{H3} ~\textit{Small multiples out perform other techniques when there are more occurrences}. Besides the height attribute, 3D small multiples also use color scale to represent the height property in order to improve its legibility in a compact space. Horizon graph technique does not perform as well as it does in the 2D visualization comparing to the small multiple technique. 
\end{itemize} 

\subsection{Visualization Tasks}

There are many tasks related to assessing graph representations. However, our hypotheses are based on the premise that each graph visualization technique has different strengths and weaknesses for different tasks. So through in-depth discussions with water resource experts, taxonomists, and ontology researchers, we identified two primary tasks important for visual identification of saturated thickness of the Ogallala: 

\begin{itemize}
\item ~\textbf{Maximum}: a simple exact location comparison across all study years ~\cite{lam2007overview}. 
\item ~\textbf{Discrimination}: a dispersed location comparison between study years ~\cite{simkin1987information}.
\end{itemize} 

\subsubsection{Maximum (Exact location comparison)}
We require participants in this task to find in a study year with the highest value at a specific location. So within a single graph type, there are several study years. We defined a location A that has the same longitude and latitude in all study years and ask participants for the year with highest saturated thickness.

\subsubsection{Discrimination (Dispersed location comparison)}
This task expects users to determine which study year has highest saturated thickness value at a location specific to each study year between graph types. We randomly select a location in each study year and ask participants to find the highest saturated thickness across all study years.

\subsection{Environment and equipments}
The experiments were conducted in a monitored laboratory which each user gets a short training before doing the tasks. Then we explain the tasks to users and ask him/her for a trial to get familiar with the test tool before doing actual tests. 
All experiments were conducted on a standard iMAC desktop computer 27 inches equipped with a keyboard and a mouse. The screen is set to 5120 x 2880 resolution. The test application was maximized on the screen. Participants only used the mouse during the experiments.

\subsection{Participants}
We recruited 10 subjects (5 males, 5 females), aged between 22 to 25. Participants were all volunteers and had normal or corrected normal vision with no color blindness.
We also conduct screening to make sure students capable of using computer and have some graph experience. For these reasons, sophomore students or above who had taken courses that have graphical representation get selected.

\subsection{Trial condition}
Our one shot trial test has a monitored number of simultaneous study years displaying on the screen. There is only a single graph type representing study years on each screen to avoid confusion to participants. With simple surface graph, we represent each surface with its own color. The other graph types have color scale to represent the height of saturated thickness. The darker the color is the higher value of saturated thickness we observe.

All trials are run with full window screen with only zooming and rotating interaction. These factors are used to allow user to rotate and understand other faces of the graphs or look into details on zooming experience. To answer any test questions, there will be a form with radio buttons for selection of correct answer(s) and a confirmation before the answer data transferred to our recording system.

\subsection{Procedure}
First we train participants with explanation of the surface graphs, small multiples and 3D horizon graphs. We also introduce simple interaction with graphs such as zooming and rotating to view different faces of the graph. 

After the training phase, participants are required to do practice test to make sure the test application is understood, and become familiar with the graphs. They are also required to explain concepts of each graph type to confirm theoretical and practical understanding. In addition, we introduce the concept of saturated thickness, and how it is described in the graphs. Participants sit in front of the monitor at comfortable distance (around 50cm) to them. It is possible to ask questions during the training phase but it is not allowed for the real trials.

Participants are required to finish all trials for a particular graph type before moving to other ones. They are instructed to complete the trials as quickly as possible. To avoid clicking mistakes, there is always a confirmation button for selection before moving to the next question.

\subsection{Study Design}
The following factors are included in our study:
\begin{itemize}
\item \textbf{Visualization type (V)}: Simple surface graph, small multiple and 3D horizon.
\item \textbf{Number of simultaneous displays (N)}: 2, 3 and 4 years.
\item \textbf{Task (T)}: Maximum and Discrimination.
\end{itemize} 

From these factors, we have \textit{V x N x S} or \textit{3 X 3 X 2 = 18} combinations for trial conditions. Each condition is repeated twice so we have a total of ~\textbf{36} trials per participant. The tasks are ordered from simple to complex for user to be prepared at the end. The order of graph types and order of number of study years are random.

There are ~\textbf{10} participants and each has ~\textbf{36} trials. Hence, the study system collects for a total of ~\textbf{10 x 36 = 360} trials for entire experiment.

\section{Results And Explanation}
\label{sec:results}

We consider accuracy as the key factor for evaluation. Since it does not matter if participants perform extremely fast within a given visualization, but in the end of the test session, the answer is incorrect. We will evaluate overall task completion accuracy and time to see the impact of number of study years. Then, we will look into comparison about accuracy and completion time of two tasks to see how graph type affects perception.

\subsection{Average completion accuracy and time}
\begin{figure}[!h]
 \centering
 \includegraphics[width=1\linewidth]{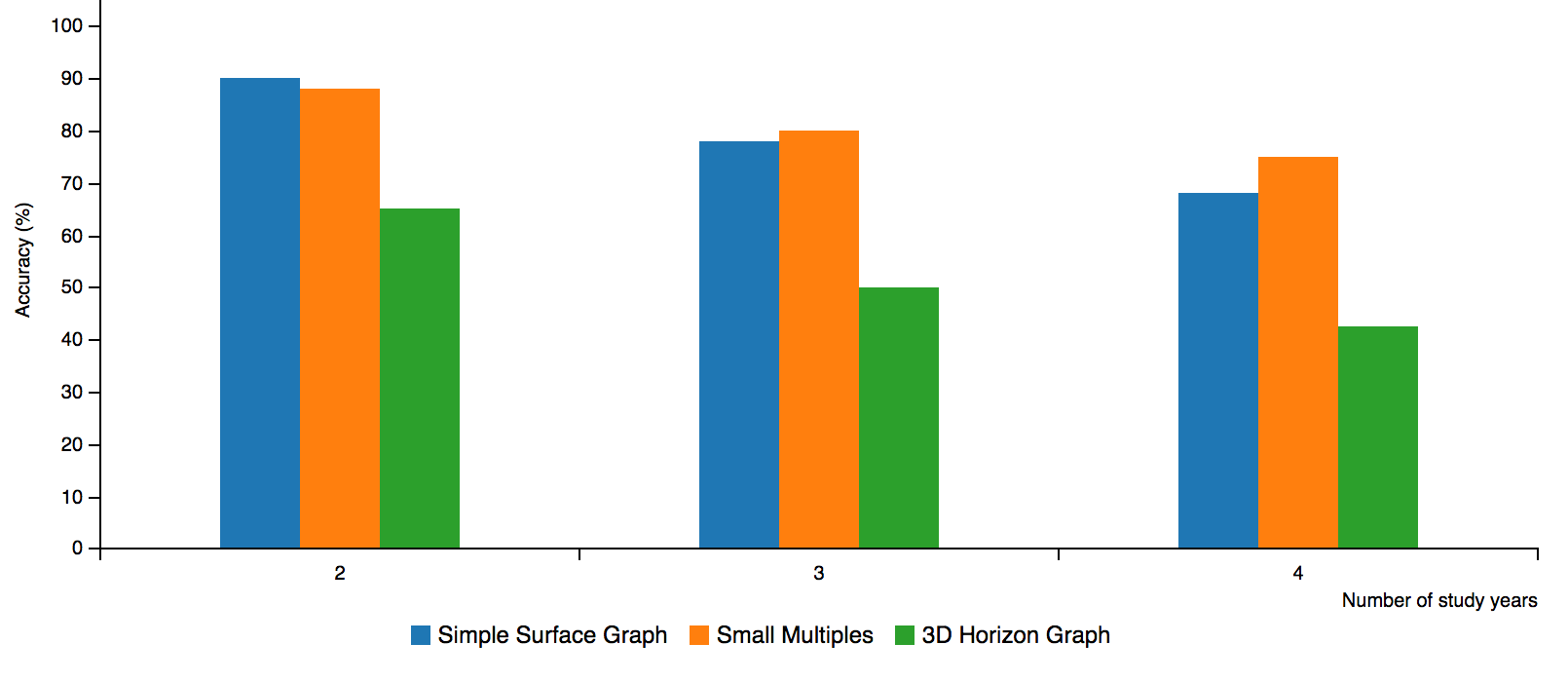}
 \caption{\label{fig:percentage-accuracy} Percentage accuracy (in \%) of tasks by number of study years}
 \vspace{-1em}
\end{figure}

From Figure \ref{fig:percentage-accuracy}, we can confirm our hypothesis ~\textbf{H1} again: the simple surface graph is more accurate with small number of occurrences (two occurrences) and it dramatically drops by number of study years from 90\% to 68\% at four occurrences. The 3D horizon graph always has lowest accuracy rate. Its performance drops quickly from 65\% at two occurrences to 42\% at four occurrences. On the other hand, the small multiple graphs slightly decrease percentage of accuracy. The change of percentage is not significant so we can tell that the small multiple technique stably maintains visual effect compared to other techniques when number of study years increases. In particular, the small multiple technique starts to get more accurate from 3 occurrences compared to normal surface graph (hypothesis ~\textbf{H3}). However, the accuracy is deducted with all graph types (hypothesis ~\textbf{H2}).

\begin{figure}[!h]
 \centering
 \includegraphics[width=1\linewidth]{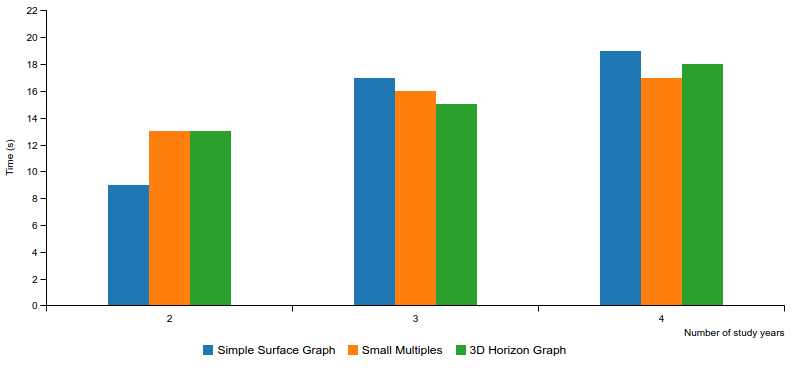}
 \caption{\label{fig:average-time} Average completion time of tasks by number of study years}
  \vspace{-2em}
\end{figure}

Below, we plot the average completion time of all tasks from our results. Figure \ref{fig:average-time} indicates that when number of study years increases, it requires more processing time in perception in order to do the task. It doubles the perception time in case of simple surface graph, from 09 seconds with two study years until around 19 seconds to complete the same task with four years representation. In contrast, the small multiple and 3D horizon graph gradually increase its completion time that it can be explained by the complexity of the representation.

\subsection{Breakdown of completion accuracy by tasks}

To understand how graph types make a different in performing tasks (simple and complex tasks), we visualize the relationship between accuracy difference between the two tasks (~\textbf{T1 - T2}) and number of study years as line graph. Each line stands for a graph type and its height is the accuracy difference. If the performance gap keeps increasing by study year, it means that the visualization technique has poor representation for number of occurrences. 

\begin{figure}[!h]
 \centering
 \includegraphics[width=1\linewidth]{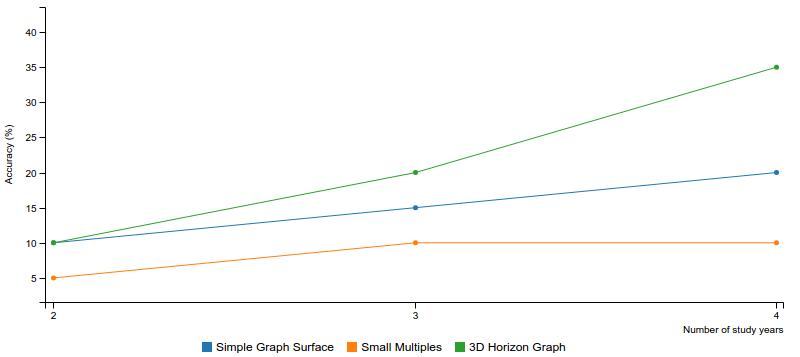}
 \caption{\label{fig:accuracy-difference-by-study-years} Accuracy difference between tasks (~\textbf{T1 - T2}) by study years}
  \vspace{-1em}
\end{figure}

From Figure ~\ref{fig:accuracy-difference-by-study-years}, we found that task ~\textbf{T1} is always more accurate than task ~\textbf{T2} across all graph types (Because the points are all positive). The 3D horizon graph always has a higher gap and increases quickly from 10\% with two occurrences to 35\% differences with four occurrences. The small multiple slightly increases and keeps stable between three and four occurrences.

From these analyses, we can see that graph type does affect user perception. It helps users to capture information in different ways so that the accuracy and time does vary with each representation. Depending on the need, 3D graphic designers can decide the graph type to visualize the information. 


\section{Conclusion And Future Work}
In this paper, we present the results of a user study approach in understanding perception regarding surface graph representations in 3D space. The results show that a few number of simple surface graphs is best for the maximum task however it shows huge clutter or overlapping resulting in information hiding with more occurrences in the graphical space. Moreover, the space that is required to visualize the graph is proportional to the saturated thickness of the graphs. The small multiple technique outperforms for larger numbers of study years. Even though the 3D horizon graph has the most compact space, its accuracy is still bottleneck.

It is obvious that the 3D representation looks more elegant compare to the 2D representation. Therefore, we are excited about our future work, which is to improve the performance of the 3D horizon graph by providing more interaction such as a slider to look into any slice of the graph mesh. Each slice represents the projection of the graph cut onto a plane. We will then investigate if this additional view will bring better information capturing for user. In addition, we will investigate more formal validation methods with support of statistical testing to ensure the correctness of our methodology.

\bibliographystyle{IEEEtran}
\bibliography{bibtemplate_samples}

\end{document}